\documentclass[showpacs,preprintnumbers,amsmath,amssymb,prb,twocolumn]{revtex4}

\def\XXint#1#2#3{{\setbox0=\hbox{$#1{#2#3}{\int}$}
     \vcenter{\hbox{$#2#3$}}\kern-.5\wd0}}

\usepackage{graphicx}
\usepackage{subfigure}
\usepackage{dcolumn}
\usepackage{bm}

\begin{document}

\title{Self-consistent T-matrix approach to Bose-glass in one dimension}

\author{A.\ G.\ Yashenkin$^{1,2}$}
\author{O.\ I.\ Utesov$^1$}
\email{utiosov@gmail.com}
\author{A.\ V.\ Sizanov$^{1,2}$}
\email{alexey.sizanov@gmail.com}
\author{A.\ V.\ Syromyatnikov$^{1,2}$}
\email{asyromyatnikov@yandex.ru}
\affiliation{$^1$Petersburg Nuclear Physics Institute NRC "Kurchatov Institute", Gatchina, St.\ Petersburg 188300, Russia}
\affiliation{$^2$Department of Physics, Saint Petersburg State University, St.\ Petersburg 198504, Russia}

\date{\today}

\begin{abstract}

Based on self-consistent T-matrix approximation (SCTMA), the Mott insulator - Bose-glass phase transition of one-dimensional noninteracting bosons subject to binary disorder is considered. The results obtained differ essentially from the conventional case of box distribution of the disorder. The Mott insulator - Bose-glass transition is found to exist at arbitrary strength of the impurities. The single particle density of states is calculated within the frame of SCTMA, numerically, and (for infinite disorder strength) analytically. A good agreement is reported among all three methods. We speculate that certain types of the interaction may lead to the Bose-glass - superfluid transition absent in our theory.

\end{abstract}


\maketitle

\section{Introduction}
	
The effect of disorder on various types of long range ordering is probably one of the most fascinating
yet not completely resolved problems in condensed matter physics. While previously this issue has been mostly addressed for fermionic systems (Ref.~\cite{and}; for recent achievements see Ref.~\cite{Evers}), the recent upsurge of interest to dirty bosons is motivated  by their novel experimental realizations in quantum simulators (ultracold atoms in optical lattices with controlled disorder, see, e.g., Ref.~\cite{Sanchez}) as well as in quantum magnets with bond disorder (for review, see Ref.~\cite{Zhel13}). The latter phenomenon is realized by substituting different atoms on peripheral sites involved into superexchange interaction paths.

From theoretical side, the picture of a Mott insulator (MI) to superfluid (SF) phase transition in presence of disorder in bosonic systems via specific gapless Bose-glass (BG) phase
originates from pioneering paper \cite{Fisher} by M.~Fisher and co-workers,
\footnote{
The absence of a direct MI$\leftrightarrow$SF transition has been proven recently as an exact theorem. \cite{theorem} For Renorm Group arguments see Refs.\cite{Kruger1, Kruger2}
}
the MI$\leftrightarrow$BG transition was found to be of Griffiths type \cite{grif1,grif}.
Furthermore, for quite general but nevertheless non-universal
types of disorder (bounded box distribution and the Gaussian unbounded one) it was demonstrated the lack of MI phase at sufficiently strong disorder or which is the same at sufficiently weak interaction \cite{Fisher}.

The physical picture behind these predictions is clear and solid because the absence of gapped phase arises from bosonic nature of excitations which therefore may condense in huge amount in rare areas of (local) random potential minima; the boson-boson repulsion prevents them from concentrating in these minima.

Generally, details of dirty boson transitions depend on combined effect of disorder and interaction which makes them quite difficult to analyze. One way is to consider weak disorder on a top of interacting pure model (for one-dimensional case see, e.g., Refs.~\cite{giam88,giam2014}). Another approach
is weakly interacting bosons subject to strong disorder \cite{Fontan, Lugan}. The methods used in the latter case were either numerical calculations \cite{Fontan} or perturbation expansion in disorder strength \cite{Lugan}.

While the disorder distributions regarded in Ref.\cite{Fisher} and in the most part of later papers
on this issue seem to us adequate for description of realistic experimental
realization of Bose condensate in optical traps and similar systems (however, cf. \cite{Krutit}), disordered quantum magnets are quite a
different matter. Indeed, the random substitution of atoms in particular sites of magnetic lattice by
another kind of atoms is definitely better modeled by binary distribution rather than by the box one.
It was the reason for us to address in the present paper the one-dimensional Bose gas subject to strong binary disorder.

\begin{figure}
 \includegraphics[scale=0.6]{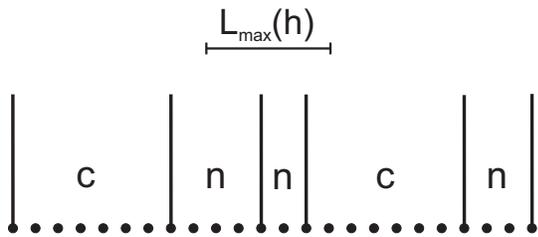}
  \caption{Particular realization of binary disorder. The peaces of the chain shorter than $L_{max}(h)$ are in the normal (gapped) state whereas the longer ones are already condensed.
  \label{forest}}
\end{figure}

We observe the physical picture of the binary case to be essentially different from the widely investigated box one. Qualitatively, it can be understood as follows. In the former case, disorder in the form of a random set of narrow equal-height potential barriers sampled over the bosonic chain splits it onto a set of finite-size {\it clean} pieces. With all the reservations caused by one-dimensionality as well as by the finite size of each piece of a chain, these pieces experience a phenomenon of "Bose condensation" at different values of a driving parameter (magnetic field) for different
pieces, depending on their lengths. As the interaction is not principally important for these transitions, we assume it to be the smallest parameter of our theory. Thus, the MI phase always exists for binary
disorder, and the BG phase is realized as a mixture of long condensed and short uncondensed "clean"
pieces (see Fig.~\ref{forest}). Furthermore, if the disorder strength is infinite (unitary limit), then there is no cross-talking of SF order parameters in different pieces, and the fully coherent SF phase never appears. At last, for noninteracting
bosons subject to strong but finite disorder we observe the picture qualitatively similar to the unitary one.

Our analysis is performed using the self-consistent T-matrix approximation (SCTMA) which allows us to go
beyond the first order in impurity concentration. This method has been successfully used for many condensed matter systems \cite{lee,ost,boson1,boson2,boson3} as a crude starting approximation.
\footnote{Typically, the SCTMA result is a subject of further corrections \cite{andrey}.}
Numerical calculations and the exact derivation of density of states (DOS) in the unitary limit provide us with  independent check of our SCTMA results.

The rest of the present paper is organized as follows. In  Sec.~\ref{GInfo} we
introduce the relevant spin models of disordered quantum magnets.
In Sec.~\ref{Theory} we describe the SCTMA technique, derive our SCTMA results for
generalized complex inverse correlation length $\varkappa$, and discuss the MI$\leftrightarrow$BG transition. In Sec.~\ref{LandDOS} we justify the analysis of BG phase, present calculations of DOS and compare the results of analytical and numerical approaches. At last, in Sec.~\ref{Summary} we summarize and briefly discuss our main results, their consequences and possible extensions.

\section{Spin Models}
\label{GInfo}

In this section we sketch two models of disordered quantum magnets. In a pure case, the lower band  excitations of both these models experience magnetic field driven "Bose condensation". When introducing
disorder, the BG phase comes into play. \cite{Zhel13}

(i) In pure case, the {\it spin-$1/2$ dimer} Hamiltonian is given by
\begin{equation}
  \begin{split}
    \mathcal{H}_{\mathbf{D}}=&\sum_i {\cal J} {\mathbf S}_{i,1} \cdot \mathbf{S}_{i,2} + \sum_{\langle i,j\rangle}  J_{ij} \left( \mathbf{S}_{i,1} \cdot \mathbf{S}_{j,1} + \mathbf{S}_{i,2} \cdot \mathbf{S}_{j,2}\right) -\\
    &- g  \mu  H \, \sum_i \left(S^z_{i,1}+S^z_{i,2} \right),
  \end{split}
  \label{ham1}
\end{equation}
where $\mathbf{S}_{i,n}$ stand for $\it n$-th spin in {\it i}-th dimer, $g  \mu  H$ is Zeeman energy in the magnetic field $H$, intra- and inter-dimer exchange couplings are denoted by $\cal J$ and $J$, respectively, and the double sums $\langle i,j \rangle$ run over nearest neighbor dimers.

The low field ground state of Eq.~\eqref{ham1} is the product of singlets on each dimer. The elementary excitations are {\it triplons}, the band of triplons being separated from the ground state by a gap. Boson version of the spin Hamiltonian \eqref{ham1} could be derived conventionally \cite{sach} by introducing three Bose operators associated with three triplet states. When applying external magnetic field $H$  triplon band  splits into three branches. For the lowest one, the gap decreases with increasing magnetic field:
\begin{equation}
  \varepsilon_{\mathbf{k}}={\cal J} - g \mu H + \frac{J_{\mathbf{k}}}{2}.
\end{equation}

(ii) The Hamiltonian of the pure {\it integer-spin chain with large single-ion easy-plane anisotropy} has the form
\begin{equation}
  \mathcal{H}_{\mathbf{A}}=\sum_{\langle i,j\rangle} J_{ij} \, \mathbf{S}_i \cdot \mathbf{S}_j + {\cal D} \sum_{i} (S^z_i)^2 -  g \mu H \, \sum_i S^z_{i},
  \label{ham2}
\end{equation}
 where the anisotropy constant ${\cal D}>0$ is assumed to be large, ${\cal D} \gg |J_{ij}|$. At low fields this Hamiltonian possesses paramagnetic ground state with $S^z_i=0$. In order to generate the Bose version of the Hamiltonian \eqref{ham2} one could introduce two types of quasiparticles (see, e.g., Ref.~\cite{Sizanov}). The lower branch of the quasiparticle spectrum reads
\begin{equation}
  \varepsilon_{\mathbf{k}}=D-g \mu H +\frac{S(S+1)}{2}J_{\mathbf{k}}.
\end{equation}
Thus, we observe that the spectrum of elementary excitations near its minimum coincides for both these models:
\begin{equation}
  \varepsilon_{\mathbf{k}}=\Delta \, + \, J  \kappa^2,
\end{equation}
where $\kappa$ is the momentum deviation from its value at the spectrum minimum. The gap $\Delta = \Delta (H) $ is found to be
magnetic field dependent. Moreover, it may vanish at large enough magnetic fields thus providing the transition to a SF state.

(iii) There are several ways to incorporate {\it disorder} into magnetic models described in (i) and (ii) \cite{Zhel13,OurPRB}. The most realistic one is to create disorder on peripheral sites involved in superexchange interaction paths thus generating the {\it bond disorder} of the form
\begin{equation}
\label{v1}
  V_{\mathbf{D}} = \sum_{\{n\}} u \, \mathbf{S}_{n,1} \cdot \mathbf{S}_{n,2},
	\text{ and }
	V_{\mathbf{A}} = \sum_{\{n\}} u \, \left(S^z_{n}\right)^2,
\end{equation}
for Hamiltonians  \eqref{ham1} and \eqref{ham2}, respectively (we assume imperfections in $\cal J$ and in $\cal D$ terms only), and the summation in Eq.~\eqref{v1} runs over all the defect sites. The bosonic versions of both models coincide:
\begin{equation}
 V_{\mathbf{D,A}} = u \, \sum_{\{n\}} a^+_n  a_n.
	\label{v}
\end{equation}

Therefore, both low energy harmonic and disordered parts of the aforementioned (physically, quite different) magnetic systems are described by the same effective Hamiltonian
\begin{eqnarray}
  \mathcal{H}&=&(\mathcal{J}-g \mu H)\, \sum_i a^+_i a_i + J \sum_i (a^+_i a_{i+1}+ a^+_{i+1} a_{i}) + \nonumber \\
  && + u \, \sum_{\{n\}} a^+_n  a_n,
  \label{BoseHub}
\end{eqnarray}
resembling disordered Bose-Hubbard model in zero interaction limit. Another peculiarity
of this problem is the negative sign of Zeeman term making
magnetic field $H$ the driving parameter for (expected) SF transition.

\section{Self Consistent T-Matrix Approximation}
\label{Theory}

In this section we introduce the formalism of SCTMA and use it in order
to investigate (possible) phase transitions from MI to BG phase as well as from BG to SF phase in
the model of disordered bosons described by Eq.~\eqref{BoseHub}. In our studies, we completely ignore the effect of quasiparticle repulsion on these transitions assuming the interaction constant to be the smallest energy scale in the problem. On the other hand, even relatively strong (but still weaker than  disorder) interaction may remain this picture undamaged for binary disorder provided the interaction simply shifts the individual transitions for all the ``clean'' pieces of the disordered bosonic chain. This scenario should be contrasted with Fisher's picture \cite{Fisher} of box distribution of disorder wherein the MI phase may completely disappear at sufficiently weak interaction. .

We shall use the conventional T-matrix approach frequently utilized for analytical description of quasiparticles subject to arbitrary disorder (see, e.g., Refs.~\cite{Izyumov,orderdif,referee1,2dvac,igar,donia}).

\subsection{Formalism. General Formulas.}

We start from single particle bosonic Green's functions in presence of disorder
\begin{equation}
  G^{R(A)} (k, E, H) =  \left( E_{\pm} - E (k) -
  \Sigma^{R(A)} (E_{\pm})     \right)^{-1},
  \label{GreenG}
\end{equation}
where $E_\pm=E\pm i 0$, "$\pm$" are referred to the retarded (advanced) Green's functions,
$E ( k) = \Delta  - g \mu H + J\,  k^2 $ is the quasiparticle spectrum, and
$\Sigma^{R(A)}$ are the quasiparticle self-energies due to disorder.
Rewriting Eq.~\eqref{GreenG} in dimensionless units we get
\begin{equation}
  g^{-1} (k, \varepsilon, h) = J^{-1} G(k, E, H) = - \left( k^2 + \varkappa^2   \right),
  \label{Greeng}
\end{equation}
where
\begin{equation}
  \varkappa^2 (\varepsilon, h) = \varkappa_{0}^{2} - \varepsilon + \sigma (\varepsilon, h),
  \label{DEFkappa}
\end{equation}
$\varepsilon = E \, / J$, $\varkappa_{0}^{2} = h_0 - h$, $h_0 = \Delta \, / J$ is the pure system critical field, $h = g \mu H \, / J$, and $\sigma = \Sigma  \, / J$. As far as the squared inverse correlation length $\varkappa^2 (h) = \varkappa^2 (0, h) $ is real and positive (weak fields) the excitations are gapped, and the system is in the MI state.
The dimensionless self-energy $\sigma (\varepsilon,h)$ is given by the following SCTMA equation:
\begin{equation}
  \sigma (\varepsilon,h)= \frac{  2 \, \alpha  \, l^{-1} }{ 1 - \alpha \,  g_0 (\varepsilon,h)},
  \label{DEFsigma}
\end{equation}
where
\begin{equation}
  g_0 (\varepsilon,h) = 2  \int \frac{d \, k}{2 \, \pi} \, g \, (k, \varepsilon,h)
  = - \frac{ {\rm sgn} \varkappa^{\prime}   }{\varkappa^{\prime} + i  \varkappa^{\prime\prime}} .
  \label{DEFg_0}
\end{equation}
 Here we allow the inverse correlation length to be complex, $\varkappa = \varkappa^{\prime} + i  \varkappa^{\prime\prime}$, $l^{-1}=c$ is the one-dimensional impurity concentration,  and $\alpha = u / \, 2 J$ is the dimensionless disorder strength. The r.h.s. of Eq.~\eqref{DEFg_0} is valid in the one-dimensional case only.

Due to non-analytic character of Eq.~\eqref{DEFg_0} the inverse correlation length $\varkappa$ obeys
different SCTMA equations depending on sign of $\varkappa^{\prime}$:
\begin{equation}
  \varkappa^{2}_{T_{-}} - \varkappa_{0}^{2} + \varepsilon = - \frac{2\, \alpha \, l^{-1} \, \varkappa_{T_{-}} }{\alpha  - \varkappa_{T_{-}} }, \quad \text{if} \quad \varkappa^{\prime} < 0,
  \label{SCTMA<0}
\end{equation}
and
\begin{equation}
  \varkappa^{2}_{T_{+}} - \varkappa_{0}^{2} + \varepsilon = \frac{2\, \alpha \, l^{-1} \, \varkappa_{T_{+}} }{\alpha  + \varkappa_{T_{+}} }, \quad \text{if} \quad \varkappa^{\prime} > 0.
  \label{SCTMA>0}
\end{equation}
The unitary limit of the infinite impurity strength $\alpha \to \infty$ is of particular interest.
Eqs.~\eqref{SCTMA<0} and \eqref{SCTMA>0} are essentially simplified in this case:
\begin{equation}
  \varkappa^{2}_{U_{\mp}} - \varkappa_{0}^{2} + \varepsilon = \mp 2 \,  l^{-1} \, \varkappa_{U_{\mp}}.
  \label{UL}
\end{equation}
Here "$-$" and "$+$" stand for $\varkappa^\prime<0$ and $\varkappa^\prime>0$ cases, correspondingly.

Also, it is convenient to introduce the Green's function in coordinate representation:
\begin{eqnarray}
  g (x, \varepsilon, h) & = & \int \frac{d k}{2 \pi} \, e^{i \, k \, x} \, g
  \,(k, \varepsilon, h) = - \, {\rm exp} \, [ - \, f (x,\varepsilon,h) ], \nonumber \\
  f(x,\varepsilon,h)  & = & x \, | \varkappa^{\prime} |
  + i \, x \, \varkappa^{\prime\prime} \, {\rm sgn} \varkappa^{\prime} + {\rm ln} \, \varkappa \,\, {\rm sgn} \varkappa^{\prime}
  \, - 2 \, {\rm ln}\,  2 \nonumber  \\
  \label{Greeng(x)}
\end{eqnarray}
This equation is simply related to the single particle density of states (DOS), the latter is  given by
\begin{equation}
  \label{DEFDOS}
	\rho \, (\varepsilon, h) = - \frac{{\rm Im}\, g\, (x=0, \varepsilon, h)}{\pi} = - \frac{\varkappa^{\prime\prime}\, (\varepsilon,h) \, {\rm sgn}  \varkappa^{\prime} (\varepsilon,h) }{4 \pi \, \, |\, \varkappa (\varepsilon,h) \, |^2 }.
\end{equation}

\subsection{Solving Equations}
\label{solveq}

Now we shall solve equations introduced in previous subsection. We start from simplest and asymptotic cases.

For clean bosonic system without impurities we get:
\begin{equation}
\varkappa^{2}_{C} \, (\varepsilon, h) = \varkappa_{0}^{2} - \varepsilon = h_0 - h - \varepsilon,
\label{CleanSq}
\end{equation}
\begin{equation}
  \varkappa_{C} \, (0, h) = \pm \sqrt{h_0 - h},
  \label{Clean}
\end{equation}
Eq.~\eqref{Clean} defines the inverse correlation length at $h<h_0$. It vanishes at the critical point
of MI$\leftrightarrow$SF transition, the corresponding Landau critical exponent being $\nu=1/2$.

Our second step is the non-self-consistent calculations. The corresponding  formulas can be easily obtained from Eqs.~\eqref{SCTMA<0}, \eqref{SCTMA>0} and \eqref{UL} if one replace $\varkappa$ with $\varkappa_0$ on the r.h.s. of these equations. The unitary limit at $h<h_0$ yields
\begin{equation}
  \varkappa_{u} = \pm \sqrt{\varkappa_0 \, (\varkappa_0   + 2  l^{-1} )   }.
  \label{ul}
\end{equation}
At  arbitrary impurity strength $\alpha$ and $h<h_0$ we obtain
\begin{equation}
  \varkappa_{t} = \pm \sqrt{ \frac{\varkappa_0 }{\alpha + \varkappa_0 }  \,
  \left( \varkappa_{0}^{2} + \alpha   \varkappa_0 + 2 \alpha l^{-1}     \right) }.
  \label{tma}
\end{equation}
Thus, the critical point remains the same ($h=h_0$) whereas the critical exponent $\nu$ changes its value from $1/2$ to $1/4$.

\begin{figure}
  \includegraphics[scale=0.43]{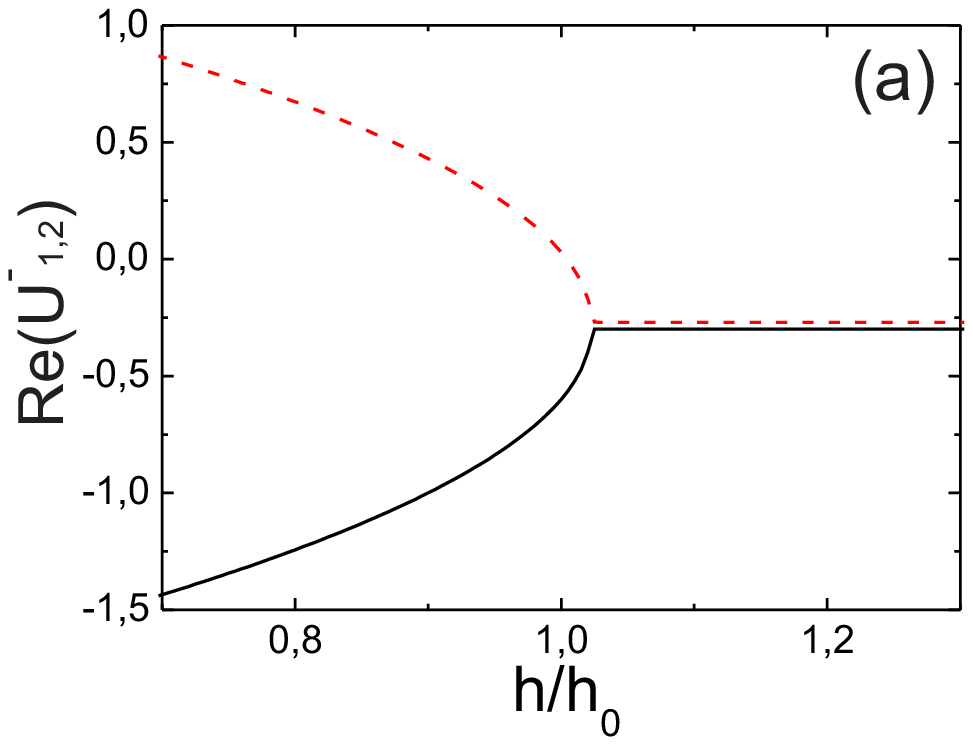}
  \includegraphics[scale=0.43]{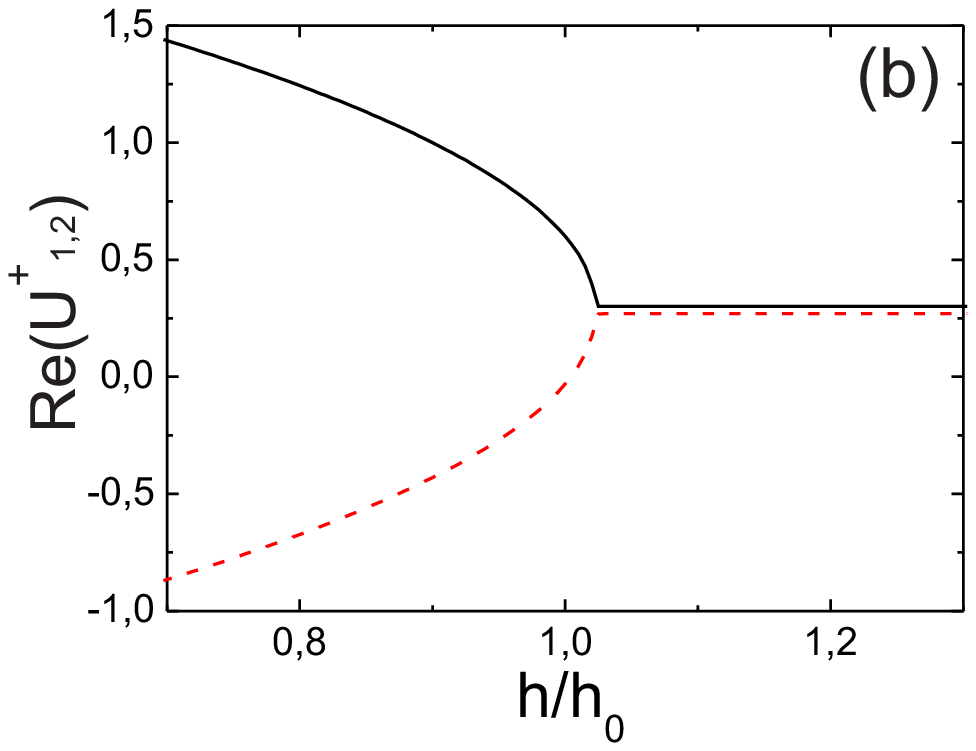}
  \caption{(Color online.) Field dependence of the unitary limit solutions ${\rm Re}U^{\mp}_{1,2}$ at $l^{-1}=0.3$. (a) $U^{-}_{1}$ and $U^{-}_{2}$ given by Eq.~\eqref{SolveU<} are shown by dashed and solid lines, respectively. (b) $U^+_{1}$ and $U^{+}_{2}$ given by Eq.~\eqref{SolveU>} are shown by solid and dashed lines, respectively.
\label{SolU}}
\end{figure}

We start our self-consistent calculations from the unitary limit. At $\varkappa^{\prime}<0$ the solutions of Eq.~\eqref{UL} are
\begin{equation}
  \label{SolveU<}
  U_{1,2}^{-}=- l^{-1} \pm \sqrt{\varkappa_{0}^{2}+l^{-2}}.
\end{equation}
One can easily see
that the only solution with negative real part is $U_{2}^{-}$. Introducing new critical field $h_1= h_0 + l^{-2}$, we  write this physical solution as
\begin{equation}
  \label{SolveU-}
  \varkappa_{U-}=-l^{-1} - \sqrt{h_1 - h},
\end{equation}
at $h < h_1 $ and as
\begin{equation}
  \label{SolveU<2}
  \varkappa_{U-}= - l^{-1} + i \sqrt{h-h_1}.
\end{equation}
at $h > h_1$. Notice that the overall sign of $\varkappa^{\prime}$ is not important since it enters
physical quantities only being squared. For $\varkappa^{\prime\prime}$ it is not the case. In particular, we choose the branch of the square root in Eq.~\eqref{SolveU<2} from the requirement of DOS to be positive.

Assuming $\varkappa^{\prime}>0$ we get
\begin{equation}
  \label{SolveU>}
  U_{1,2}^{+}=l^{-1} \pm \sqrt{\varkappa_{0}^{2}+l^{-2}},
\end{equation}
Similar to the $\varkappa^\prime < 0$ case we choose one of the solutions, namely $U^{+}_1$. It yields
\begin{equation}
  \label{SolveU+}
  \varkappa_{U_{+}}=l^{-1} + \sqrt{h_1 - h},
\end{equation}
at $h<h_1$ and
\begin{equation}
  \label{SolveU>2}
  \varkappa_{U_{+}} = l^{-1} - i \sqrt{h-h_1}.
\end{equation}
at $h>h_1$.
Solutions Eqs.~\eqref{SolveU-}--\eqref{SolveU<2} and Eqs.~\eqref{SolveU+}--\eqref{SolveU>2} differ only by the overall sign, so they are physically equivalent. Plots of the real parts of both these solutions are presented in Fig.~\ref{SolU}.
As $h$ reaches its critical value $h_1$, the real part of the inverse correlation length $\varkappa_{U_{\pm}}$ remains finite whereas the {\it imaginary} part comes into play.
One can easily check that this unambiguously results in finite DOS at zero energy (see next section). The latter phenomenon is generally associated with onset of the BG state \cite{Fisher}.
Also, it should be mentioned that at $h_2=h_1 + l^{-2}$ the real part of $\varkappa^2$ changes its sign
which reflects the fact that certain portion of chain pieces is already in the superfluid state. With further increase of the magnetic field the correlation length remains finite. It
manifests the absence of BG$\leftrightarrow$SF transition in the unitary limit. Physically, it is the obvious consequence of infinite impurity barriers separating different pieces of the superfluid phase thus preventing the global coherence of the chain.

Now we turn to arbitrary impurity strength $\alpha$. In this case, we have two different SCTMA equations \eqref{SCTMA<0} and \eqref{SCTMA>0}. They could be transformed into cubic equations for $\varkappa_{T_{\mp}}$ and solved analytically. First, we present three solutions $T^{-}_{1,2,3}$ of Eq. \eqref{SCTMA<0}:
\begin{eqnarray}
  \label{Negative}
	T^{-}_{1}& = & \frac{1}{3} \left(\, \alpha + c_0 \, A \, Q^{-1} + c_0 \, Q  \, \right),  \nonumber \\
  T^{-}_{2}& = & \frac{1}{3} \left(\, \alpha + c_1 \, A \, Q^{-1} + c_2 \, Q  \, \right),  \\
  T^{-}_{3}& = & \frac{1}{3} \left(\, \alpha + c_2 \, A \, Q^{-1} + c_1 \, Q  \, \right),  \nonumber
\end{eqnarray}
where $c_0 =1, \quad c_{1,2} = - (1 \pm i \sqrt{3})/2$, and
\begin{eqnarray}
  A& = & 3\, (h_3 - h - \varepsilon),  \nonumber \\
  h_3 & = & h_0 + \alpha^{2}/3 + 2 \, \alpha \, l^{-1}, \nonumber \\
	\label{q}
  Q& = & \left(\sqrt{B^2 - A^3} +B \right)^{1/3},  \\
	\label{b}
  B& = &  \alpha \left( \alpha^2 +  9 \, (\alpha \, l^{-1} + h-h_0+ \varepsilon)  \right).
\end{eqnarray}

\begin{figure}
  \includegraphics[scale=0.46]{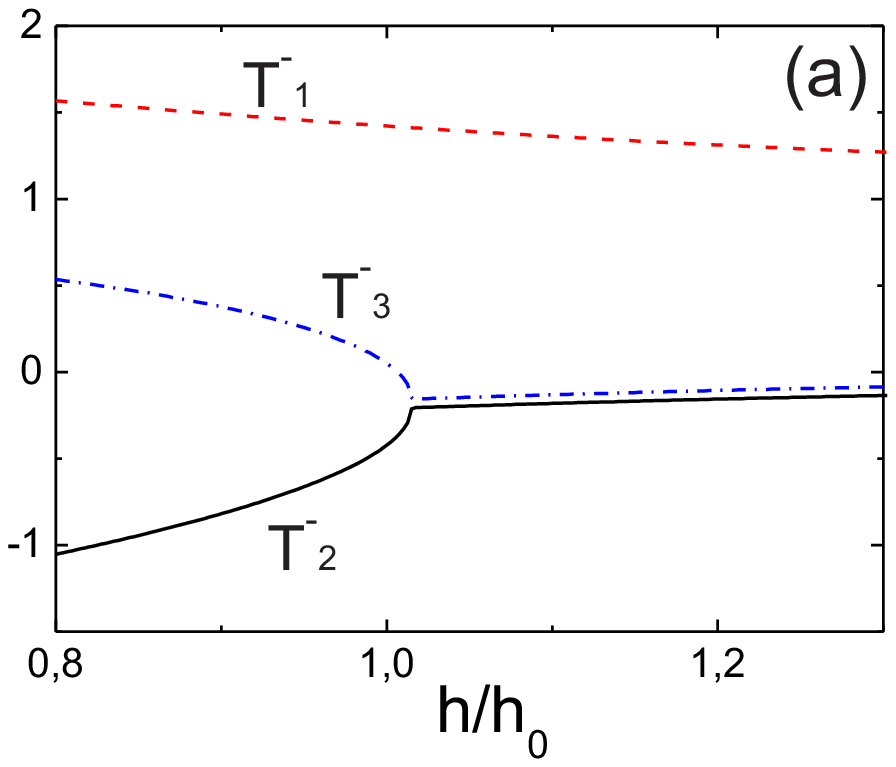}
  \includegraphics[scale=0.46]{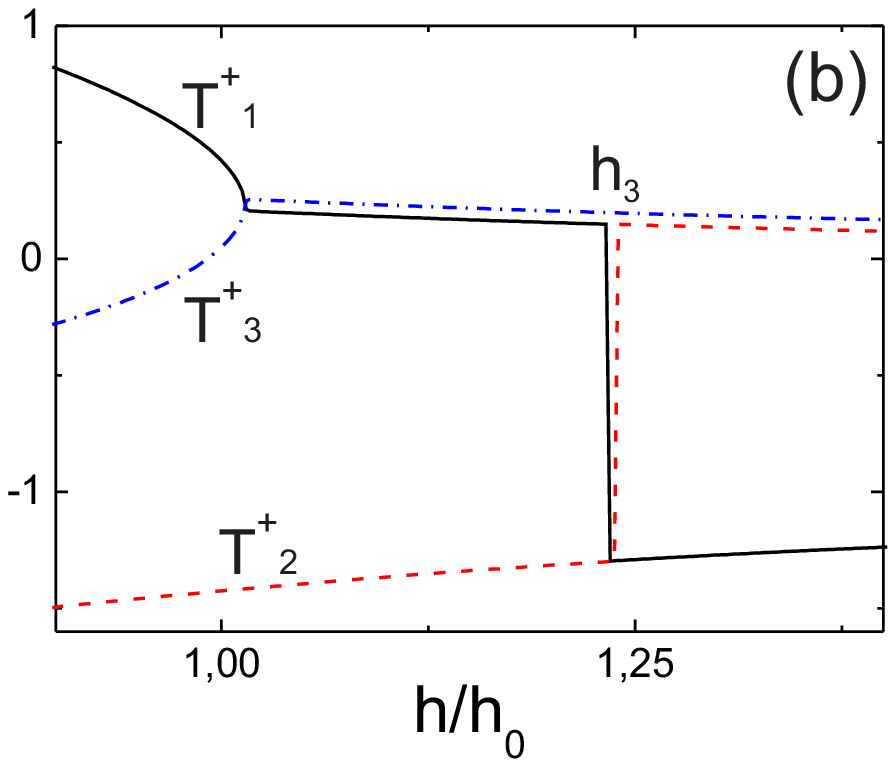}
  \caption{(Color online.) Field dependence of SCTMA solutions ${\rm Re}T^{\pm}_{1,2,3}$ at $\alpha=1$ and $l^{-1}=0.3$. (a) $T^{-}_{1,2,3}$ given by Eqs.~\eqref{Negative}. (b) $T^{+}_{1,2,3}$ given by Eqs.~\eqref{Positive}.
	\label{SolT}}
\end{figure}

Real parts of all these solutions are depicted in Fig. \ref{SolT}(a). The only one which is negative below the critical field is $T^{-}_{2}$. Furthermore, one can readily check that only this solution $T^{-}_{2}$ reaches the unitary one $\varkappa_{U-}$ as the impurity strength $\alpha$ increases. Therefore, we should consider $T^{-}_{2}$ as the physical solution:
\begin{equation}
  \varkappa_{T_{-}} (\varepsilon,h, \alpha, l )= T^{-}_{2} (\varepsilon,h, \alpha, l ).
  \label{SolT<}
\end{equation}

For $\varkappa^{\prime}<0$ case, we obtain the following scenario. At critical field $h_{1} (\alpha)$ the system possesses transition from MI to BG phase, $h_0 < h_{1} (\alpha) < h_1$. Furthermore, the correlation length remains finite in higher fields $h> h_{1} (\alpha)$ similarly to the unitary case demonstrating the absence of the BG$\leftrightarrow$SF transition. On the other hand, the finite DOS at zero energy appears starting at $h = h_{1} (\alpha) $ as a manifestation of the MI$\leftrightarrow$BG transition. In general, the picture of the inverse correlation length dependence on the magnetic field remains qualitatively the same for all values of the disorder strength $\alpha$. Remember that
\begin{eqnarray}
h_1(\alpha\to0) &=& h_0,\\
h_1(\alpha\to\infty) &=& h_1.
\end{eqnarray}

The situation looks quite different when we are considering the solutions $T^{+}_{i}$ ($i=1,2,3$) of SCTMA equation at
$\varkappa^{\prime}>0$, see Eq.~\eqref{SCTMA>0}. They could be easily found by noticing that
\begin{equation}
  T^{+}_{i} (\varepsilon,h, \alpha, l )=T^{-}_{i} (\varepsilon,h, -\alpha, -l ).
  \label{Positive}
\end{equation}
Performing the analysis similar to that carried above (see Fig.~\ref{SolT} (b)) we choose as the physical
 solution at $h < h_3$
\begin{equation}
  \varkappa_{T_{+}} (\varepsilon,h, \alpha, l )= T^{+}_{1} (\varepsilon,h, \alpha, l ).
  \label{SolT>}
\end{equation}
However, at $h > h_3$ this solution experiences a jump to a large negative value thus violating self-consistency. Mathematically, this difference with $\varkappa^{\prime}<0$ case stems from the fact that at $\varkappa^{\prime}>0$ the sign in front of $B$ in Eq.~\eqref{b} is opposite and thus the expression under the cubic root in Eq.~\eqref{q} for $Q$ could change the sign. Explicitly, the following factor arises in $B$ at $h\approx h_3$:
\begin{equation}
  \frac{h_3 - h}{\left( (h_3 - h)^3   \right)^{1/3}}, \nonumber
  \label{Jump}
\end{equation}
which gives the jump at $h = h_3$ {\it if we take the same definite branch of $(-1)^{1/3}$ for all values of the magnetic field}.

\begin{figure}
  \includegraphics[scale=0.7]{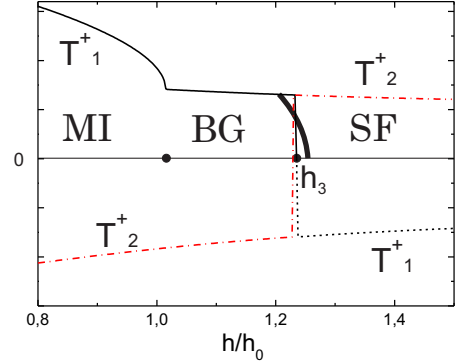}
  \caption{(Color online.) Solutions of Eq.~\eqref{SCTMA>0}. Thick black curve illustrates what can happen with the jump if interaction comes into play.
	\label{SolJumps}}
\end{figure}

Thus, we arrive at the following dichotomy considering the solution of SCTMA equation for $\varkappa^{\prime}>0$. According to Fig.~\ref{SolJumps}, the system can start to follow
another solution at $h =h_3$ ($T^{+}_{2}$ rather than $T^+_1$) and, in comparison with $\varkappa^{\prime}<0$ case, nothing physically new will happen. Alternatively, the system may follow $T^{+}_{1}$ solution all the time and particularly through $h=h_3$ point where both real and imaginary parts of $\varkappa$ simultaneously jump to zero (see Figs.~\ref{SolT}(b) and \ref{SolJumps}). It would manifest a transition from BG phase into another one. Our numerics
for non-interacting bosons presented in next section demonstrates that the former scenario realizes. However, we believe that still there remains a possibility to stabilize the $\varkappa^{\prime}>0$ solution by including certain type of boson-boson
interaction. It would transform the jump to some critical curve as it shown in Fig.~\ref{SolJumps} by the thick black line separating BG and SF phases. This is the problem we not address in the present paper.

\begin{figure}
  \includegraphics[scale=0.43]{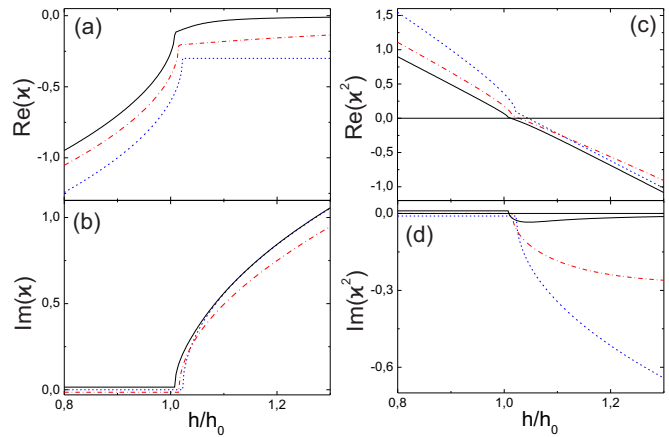}
  \caption{(Color online.) a) and b) The SCTMA physical solutions $\varkappa_{T_{-}}(h)$ for $\varkappa'<0$, $\varepsilon=0$, $l^{-1}=0.3$ and different $\alpha$. Solid black, dashed-dotted red, and dotted blue curves are for $\alpha=0.2$, $\alpha=1$, and for $\alpha\to\infty$, respectively. c), d) The same for $(\varkappa_{T_{-}}(h))^2$.  }
  \label{Sol-}
\end{figure}

\begin{figure}
  \includegraphics[scale=0.43]{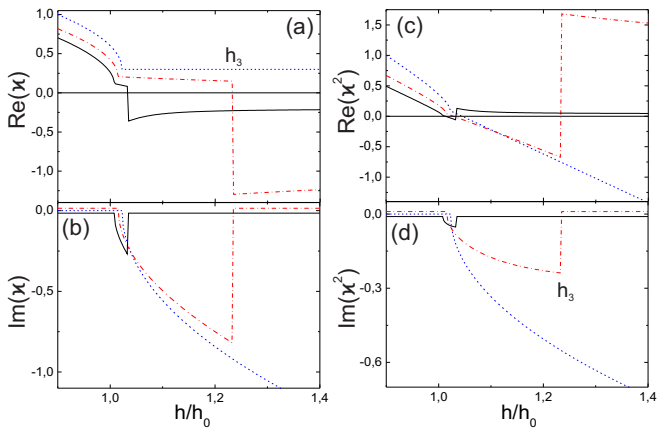}
  \caption{(Color online.) Same as in Fig.~\ref{Sol-} for $\varkappa_{T_{+}}$.
  \label{Sol+} }
\end{figure}

In Figs.~\ref{Sol-} and \ref{Sol+} we plot both real and imaginary parts of the inverse correlation length for $\varkappa^{\prime}<0$ and $\varkappa^{\prime}>0$ cases, respectively, as functions of magnetic field for different values of the impurity strength $\alpha$. We see that in both cases the qualitative behavior remains quite the same when $\alpha$ changes from small values to infinity.
On the other hand, all the "critical" values of the external magnetic field are $\alpha$ dependent when the solutions qualitatively change their behavior.

\section{Gap and density of states}
\label{LandDOS}

There are several problems to be addressed in this section. First, we study the SCTMA solutions for generalized complex inverse correlation length $\varkappa (\varepsilon, h)$ derived in previous section in order to establish its
physical meaning, particularly in the field range of ``negative gap''. Second, we present the SCTMA analytical results for the density of  single particle states of disordered bosons. Third, we derive analytically the exact DOS in the unitary limit. At last, we check and verify our analytical approaches by presenting numerics for DOS of noninteracting dirty bosons.

\subsection{Negative Gap}

The SCTMA analytical expressions for $\varkappa (\varepsilon, h)$ obtained above reveal the following general
structure common for unitary limit and finite $\alpha$ case. At low fields $h< h_1$ the MI state is
characterized by a finite real positive spectrum gap $\varkappa^2$. Furthermore, at $h=h_1(\alpha)$ the inverse
correlation length becomes complex which manifests the appearance of the BG phase, but the real part
of $\varkappa^2$ still remains positive although decreases. At last, in higher fields $\varkappa^2$ changes its sign to the negative one. The question naturally arising is: does the SCTMA description in this field interval make any sense?

\begin{figure}
  \includegraphics[scale=0.3]{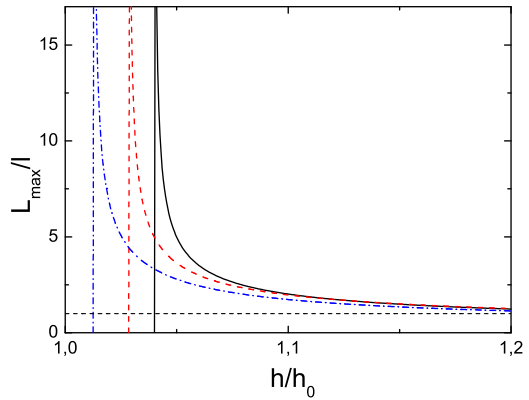}
  \caption{(Color online.) ${\cal L}_{max}(h)$ for various impurity strengths $\alpha$ and $l^{-1}=0.3$. Solid black, dashed red, and  dashed-dotted blue curves corresponds to $\alpha=5$, $\alpha=1$, and $\alpha=0.2$, respectively.
	\label{Lmax}}
\end{figure}

Our positive answer to the above question is based on the following qualitative picture of transitions in the disordered bosonic chain {\it with binary distributed disorder} already mentioned in the introduction.
Indeed, this type of disorder introduces the randomly sampled equal-height potential barriers into the
chain thus splitting it onto a set of finite size $\it clean$ pieces of different lengths.
Because of their finiteness, they experiences the one-dimensional finite-size analog of Bose condensation at {\it various  values} of the external magnetic field: the longer particular piece of the chain is, the lower field
is needed to move this piece into SF state. Thus, the BG phase could be considered as a mixture of already
condensed in this field longer pieces of the bosonic chain and the shorter ones still remaining in the gapped state. The derivation of the disorder averaged two-component single-particle Green's function in the BG phase is  quite a difficult task. The
only what we know about it from general arguments is that it should be gapless. However, we
can "regard" this mixed state from the MI side noticing that the condition ${\rm Re} E (k) > 0 $
at negative values of the gap provides us with certain limitation on the momenta (and therefore on lengths) which could be consistently treated within this approach. We interpret this limiting length
${\cal L}_{max} (h)$ as the maximal size of pieces which still remain in insulating state at a given magnetic field. More specifically, this quantity is defined by the following equation:
\begin{equation}
  {\cal L}_{max} (h,\alpha, l)  = \pi \left(- \varkappa^{2 \, \prime} \, (h, \alpha, l)   \right)^{-1/2}
  \label{length}
\end{equation}
This formula is illustrated in Fig.~\ref{Lmax} at various values of the parameter $\alpha$.

Two obstacles should be mentioned here. First, there is a gap in the low field dependence of ${\cal L}_{max}$ in Fig.~\ref{Lmax}, so it looks like there is no condensed
pieces within this interval. Below we shall argue that it is an artifact of  the SCTMA which is too crude to reproduce the exponential tails in the vicinity of the real value $h=h_0$ of the critical magnetic field for the MI$\leftrightarrow$BG transition.

Our second remark is about the possibility of BG$\leftrightarrow$SF transition. As we already mentioned above,
even when all the pieces of the chain already experience the (local) Bose condensation it does not necessarily lead to the onset of the global superfluid coherence throughout the entire chain. This coherence
definitely never appears in the unitary limit where the infinite heights of the impurity potentials do not allow different pieces of the chain to cross-talk at any given magnetic field.

\subsection{Evaluating DOS}

\begin{figure}
  \includegraphics[scale=0.97]{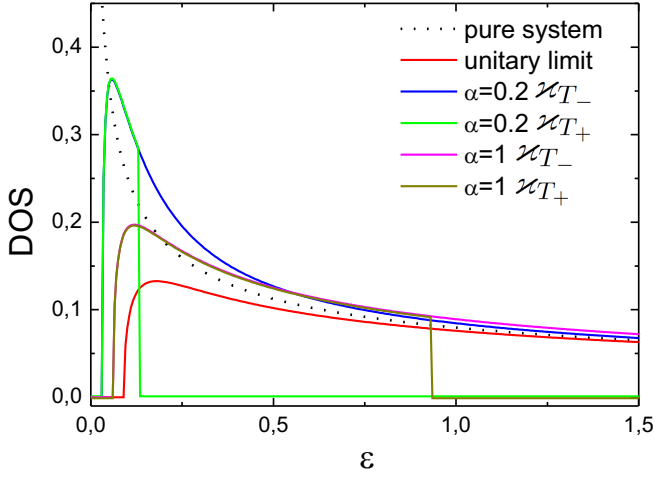}
  \caption{(Color online.) DOS at $h=h_0$ and $l^{-1}=0.3$.
	\label{T-dos}}
\end{figure}

The density of states obtained within the framework of the SCTMA is given by general Eq.~\eqref{DEFDOS}. As we already know all quantities entering this expression we present the energy dependence of DOS in Fig.~\ref{T-dos} for several values of $\alpha$.

Yet another more direct and approximation free method to calculate DOS exists in the unitary limit. Remind that in this limit our system is decomposed onto many pieces isolated from each other. Energy levels for a given chain of length $L$ (dimensionless units) is
\begin{equation}
  \varepsilon_m(L)=\varkappa^2_0+2-2 \cos{\frac{\pi m}{L+1}}.
  \label{SWenergy}
\end{equation}
Furthermore, in the chain of total length $N$ the probability to find a pure sub-chain of length $L$ can be written as follows
\begin{equation}
  p(L)=c^2 (1-c)^L=c^2 e^{-\beta L}, \quad\beta=-\log{(1-c)},
\end{equation}
where $c=l^{-1}$ is  the impurity concentration. For the number of such chains we get
\begin{equation}
  n(L)= N p(L)=c^2 N e^{-\beta L}.
  \label{Nchain}
\end{equation}
The DOS is defined by the equation
\begin{equation}
  \rho(\varepsilon)=\frac{1}{N}\frac{\delta M(\varepsilon, \delta \varepsilon)}{\delta \varepsilon}=\frac{1}{N} \sum_m \frac{n(L) \delta L}{ (\partial \varepsilon_m(L)/\partial L) \delta l}.
\end{equation}
Here $M(\varepsilon, \delta \varepsilon)$ is a number of states within the  energy interval  $[\varepsilon, \, \varepsilon+\delta \varepsilon]$. As we are interested only in states near the band bottom, only first few harmonics have an impact on $\rho(\varepsilon)$, and the contribution from others is exponentially negligible. So we can replace the upper limit of summation over $m$ by infinity. Introducing the function $L(\varepsilon,m)$ form Eq.~\eqref{SWenergy}
\begin{equation}
  L(\varepsilon,m) = \frac{ \pi m }{ \arccos \left[ (\varkappa_{0}^{2}-\varepsilon)/2 + 1 \right]}-1
\end{equation}
we get
\begin{equation}
  \rho(\varepsilon) = c^{2} \sum_{m=1}^{\infty} \frac{\partial L(\varepsilon,m)}{\partial \varepsilon} e^{-\beta L(\varepsilon,m)}.
\end{equation}
Explicitly:
\begin{eqnarray}
\label{rhoUL}	
\rho(\epsilon) &=& \frac{\pi c^{2} }{ 2\arccos^{2}(x) \sqrt{1-x^{2}} (1-c) }\\
 &&\times \sum_{m=1}^{\infty} m \exp \left[-\beta \frac{  \pi m }{\arccos(x)} \right], \nonumber	 \\
x &=& \frac{\varkappa_{0}^{2} -\varepsilon}{2} + 1.		
\end{eqnarray}
Performing the summation in Eq.~\eqref{rhoUL} we rewrite the density of states in the following form:
\begin{equation}
  \label{Udos}
  \rho(\epsilon) = \frac{ \pi c^{2} }{ 8(1-c) }
	\frac{1} {
		   \sinh^{2} \left[  \frac{\beta \pi }{ 2\arccos(x) } \right]
		   \arccos^{2}(x)
		   \sqrt{1-x^{2}}
		 }.
\end{equation}
Another frutful formula is the asymptote of Eq.~\eqref{Udos} near the band bottom:
\begin{eqnarray}
\label{rholim}
\rho(\epsilon \to \varkappa^{2}_{0}) \rightarrow
		   \frac{\pi c^{2} }{2(1-c) \left(\epsilon - \varkappa^{2}_{0} \right)^{3/2} }
		   \exp \left[ - \frac{ \beta \pi }{\sqrt{\epsilon - \varkappa^{2}_{0}}} \right].
\end{eqnarray}

\begin{figure}
  \includegraphics[scale=0.8]{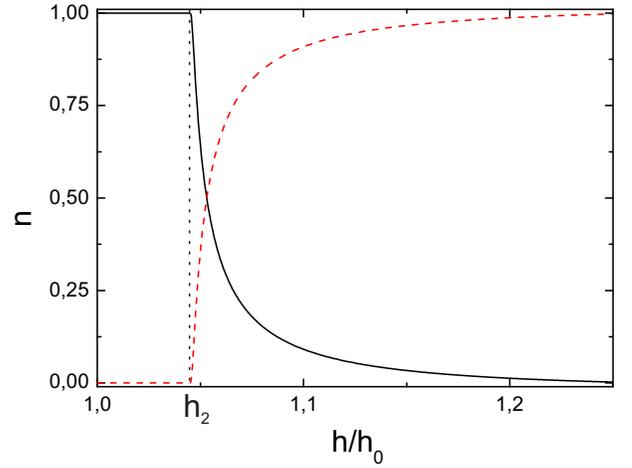}
  \caption{(Color online.) Fractions of normal $n_n$ (solid black) and condensed $n_c$ phases (dashed red) as functions of field in the unitary limit for $l^{-1}=0.3$.
  \label{fract}}
\end{figure}

Within the above approach we calculate the fractions of sites in "normal" gapped phase $n_n$ and in "condensed" phase $n_c$. Using Eq.~\eqref{length} and Eq.~\eqref{Nchain} yields
\begin{eqnarray}
  n_n&=&1-c {\cal L}_{max}(1-c)^{ {\cal L}_{max}-1}-(1-c)^{ {\cal L}_{max}}, \\
  n_c&=&c {\cal L}_{max} (1-c)^{ {\cal L}_{max}-1}+(1-c)^{ {\cal L}_{max}}.
\end{eqnarray}
This formulas for $c=l^{-1}=0.3$ are illustrated in Fig.~\ref{fract}.
Notice the fact that the changes in $n_n$ and $n_c$ in this figure start at $h=h_2$. Again, it is an artifact of our crude SCTMA Eq.~\eqref{length}. In reality, it ignores exponentially weak variations of this quantities between $h_0$ and $h_2$ (cf. discussion of ${\cal L}_{max}$ and Eq.~\eqref{rholim}).

\begin{figure*}
 \includegraphics[scale=0.4]{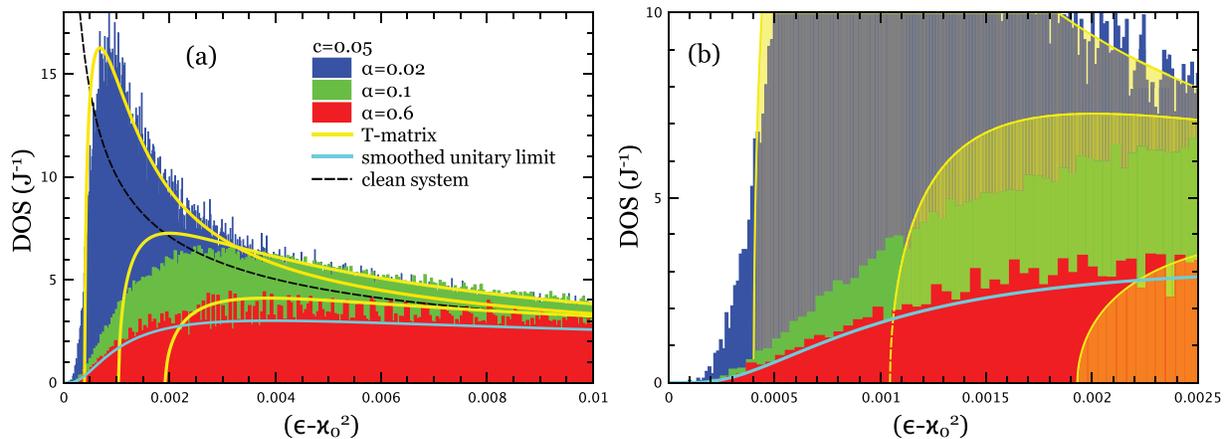}
 \caption{(Color online.) Energy dependence of DOS in the entire energy range (a) and at low energies in the "tail" area (b). Shaded areas correspond to numerics. The curve "smoothed unitary limit" is drawn using Eq.~\eqref{Udos}.
\label{dos-gr_}}
\end{figure*}

In order to check our analytical predictions and the validity of both SCTMA and the above formulas
we accompany our analytical calculations by the numerics. More specifically, the DOS was evaluated
numerically with the use of procedure of exact diagonalization of the single particle sector of  Hamiltonian \eqref{BoseHub} and further averaging of the results over the large number of realizations (several thousands). We used {\it NumPy} library for these purposes. We introduced periodic boundary conditions for the chains of lengths varying from 8000 to 12000 sites. Notice that the discrete binary on-site character of disorder made it possible to reach good accuracy including the description of exponential behavior near the bottom of the single-particle band dominated by rare regions effects.

Comparison of SCTMA DOS, exact unitary DOS given by Eq. \eqref{Udos} and numerically calculated one is presented in Fig.~\ref{dos-gr_}. We see that the accuracy of analytical approaches varies from good to very good.
In particular, all general tendencies of curve transformations with change of the parameters are
correctly reproduced. Furthermore, the formula for DOS derived in this subsection fits
the corresponding numerics excellently including the range of exponential tail in weak magnetic fields.
Simultaneously, our SCTMA curves are very good at large and intermediate energies but they all
predict sharp thresholds at lower ones while numerics and the exact unitary limit formula demonstrate smoother tail-like
behavior in this region. We conclude that the SCTMA is too crude
to reproduce the exponential tails (cf. the result of more delicate approach for the unitary
limit). Moreover,
there exist exact arguments in literature \cite{Zhel13} that disorder does not shift the critical field $h_0$ of a clean MI$\leftrightarrow$SF transition just replacing this transition by the MI$\leftrightarrow$BG one. It follows from our comparison that the SCTMA mimic the exponential tail started at $h_0$ by shifting the critical field to higher value  $h_{1} (\alpha)$.



\section{Summary and conclusions}
\label{Summary}

In our paper, we report the MI phase for arbitrary strength of binary disorder. This differs essentially from the case of disorder in the form of bounded box distribution addressed in Ref.~\cite{Fisher} and in many subsequent papers wherein the MI phase disappears for wide enough width of disorder distribution function or for weak enough interaction between the bosons. We believe that our model \cite{OurPRB} is more suitable for description of real quantum magnetic spin systems. \cite{Zhel13}

Furthermore, our approach is based on a combined use of SCTMA, numerical, and exact analytical calculations. We obtain MI$\leftrightarrow$BG transition but we are unable to reproduce the BG$\leftrightarrow$SF one. We believe that one has to take into account the interaction between the bosons in order to examine the latter transition. As soon as our main goal was to describe the MI$\leftrightarrow$BG transition, we assume the boson repulsion to be the smallest parameter of the theory which can be omitted. We believe that for certain types of interaction the second noninteracting SCTMA scenario of Sec.~\ref{solveq} which was rejected from the comparison with numerics may be reincarnated allowing for BG$\leftrightarrow$SF transition. This promising issue deserves further consideration.

It is worth mentioning that the recent paper \cite{giam2014} considered the same problem of one-dimensional dirty bosons utilized the approach solid within the domain of strong interaction/weak disorder which is essentially the complimentary domain of parameters we are dealing with.

To conclude, based on self-consistent T-matrix approximation we address the Mott insulator - Bose-glass phase transition of one-dimensional noninteracting bosons subject to binary type of disorder. We argue that this type of defects is the most suitable one to describe the recently fabricated quantum disordered magnets. We find our results to be essentially different from the widely discussed in literature case of box distribution of the disorder. In particular, we predict MI$\leftrightarrow$BG transition to exist for arbitrary strength of impurities. We calculate the single-particle density of states within the framework of SCTMA as well as numerically and (for infinite disorder strength) analytically. We observe a very good agreement between results obtained with use of all three methods. However, SCTMA being a crude approximation fails to reproduce exponential tails (e.g., in DOS) stemming from rare events. While our approach does not describe the Bose-glass - superfluid transition, we speculate that the incorporation of certain types of bosonic interaction may resolve this problem.

\begin{acknowledgments}

Authors are benefited from discussion with S.V.\ Maleyev and from correspondence with K.V.\ Krutitsky. This work is supported by Russian Scientific Fund Grant No.\ 14-22-00281. One of us (O.I.U.) acknowledges the Dynasty foundation for partial financial support. A.V.\ Sizanov acknowledges Saint-Petersburg State University for research grant 11.50.1599.2013.

\end{acknowledgments}

\bibliography{bibliography}

\end{document}